\documentclass{jetpl}
\twocolumn
\usepackage{epsfig,bm,amsfonts,amssymb}

\newcommand{\be}{\begin{eqnarray}}
\newcommand{\ee}{\end{eqnarray}}
\newcommand{\ba}{\begin{array}}
\newcommand{\ea}{\end{array}}

\lat

\title{Leading Chiral Logarithms for Pion Form Factors to Arbitrary Number of Loops}

\rtitle{Leading Chiral Logarithms for Pion Form Factors}

\sodtitle{Leading Chiral Logarithms for Pion Form Factors}

\author{Nikolai~A.~Kivel$^{1,2}$, Maxim~V.~Polyakov$^{1,2}$, Alexei~A.~Vladimirov$^{1,3}$}

\rauthor{N.~A.~Kivel, M.V.~Polyakov, A.~A.~Vladimirov}

\sodauthor{N.~A.~Kivel, M.V.~Polyakov, A.~A.~Vladimirov}

\address{$^1$ Institut f\"ur Theoretische Physik II,
Ruhr--Universit\"at Bochum, D--44780 Bochum, Germany\\
$^2$Petersburg  Nuclear Physics Institute, Gatchina, 188300, St.
Petersburg, Russia\\
$^3$ Bogolubov Laboratory of Theoretical Physics, JINR, 141980 Dubna, Russia}

\abstract{We develop the method of calculation of the leading chiral (infrared) logarithms to an arbitrary loop order for various form factors
of Nambu-Goldstone bosons. The method is illustrated on example of scalar and vector form factors in massless 4D $O(N+1)/O(N)$ $\sigma$-model.
The analytical properties of the form factors are derived. The leading chiral (infrared) logarithms are summed up in the large $N$ limit.}


\begin{document}

\maketitle

Recently we developed a new puissant method  \cite{KPVprl} for
calculations of leading chiral (infrared) logarithms in a wide class
of non-renormalizable massless field theories. The method has been
applied to the amplitude of Nambu--Goldstone boson scattering in
4D $O(N+1)/O(N)$ sigma-model defined by the following Lagrangian:

\be \label{ON} \mathcal{L}_{2}&=&\text{~}\frac{1}{2}\left[
\partial_{\mu}\sigma\partial_{\mu
}\sigma+\partial_{\mu}\pi^{a}\partial_\mu\pi^{a}\right] \, , \ee
where the fields are constrained by the relation
$\sigma^{2}+\sum_{a=1}^{N}\pi^{a}\pi^{a}=F^{2}\, .$ The amplitude
of $\pi\pi$ scattering can be decomposed into the invariant tensors of
$O(N)$ group as follows: \be \label{amp}
T^{abcd}=\delta^{ab}\delta^{cd}A+\delta^{cb}\delta^{da}B+\delta^{bd}\delta^{ac}C.
\ee The amplitudes $A,B$ and $C$ are functions of the
Mandelstam variable $s$ and c.m. scattering angle $\theta$. In
Ref.~\cite{KPVprl} we derived these amplitudes in the leading
logarithms (LLs) approximation. The LLs approximation consists in
the summation of contributions of the type $\sim \left[s \ln s
\right]^n$ in the low-energy expansion of the amplitude. Such
contributions arise from the $n$-loop Feynman graphs of the
effective theory (\ref{ON}). At the first glance, a mission is impossible
-- to compute the n-loop graphs in a non-renormalizable
theory. However, in the LLs approximation this task can be
accomplished as it can be reduced to the calculations of the spectrum of
anomalous dimensions of the $O(N)$ symmetric composite operators
made of four pion fields in an  {\it free field theory}
\cite{KPVprl}. The result Ref.~\cite{KPVprl} for the $\pi\pi$ scattering amplitudes
in LLs approximation can be presented in the form of the partial wave decomposition as:
\be
\label{AMPpipi}
A&=&\frac{s}{F^2} \sum_{n=0}^\infty \left(S\ L \right)^n \sum_{l=0\atop \scriptstyle{\rm even}}^{n+1} \omega_{nl}\ P_l(\cos\theta),\\
\nonumber
B&=&\frac{s}{F^2} \sum_{n=0}^\infty \left(S\ L \right)^n \sum_{j=0\atop \scriptstyle{\rm even}}^{n+1} \omega_{nj}\sum_{l=0}^{n+1}
\Omega_{n+1}^{jl}\ P_l(\cos\theta),\\
\nonumber
C&=&\frac{s}{F^2} \sum_{n=0}^\infty \left(S\ L \right)^n
\sum_{j=0\atop \scriptstyle{\rm even}}^{n+1} \omega_{nj}\sum_{l=0}^{n+1}
(-1)^l\Omega_{n+1}^{jl}\ P_l(\cos\theta).
\ee
We introduced dimensionless expansion parameter $S=s/(4\pi F)^2$ and we collected only contribution with maximal power of chiral logarithm $L=\ln(-\mu^2/s)$,
$P_l(\cos\theta)$ are Legendre polynomials. The crossing matrix $\Omega_n^{jl}$
has the following form:
\be
\label{omega}
\Omega_n^{jl}=\frac{2 l+1}{2^{n+1}}\int_{-1}^1 dx P_j\left(\frac{x+3}{x-1}\right) P_l(x) (x-1)^n.
\ee
The LLs coefficients $\omega_{nj}$ satisfy non-linear recursion relation \cite{KPVprl}:
\be
\label{main}
\omega_{nj}=\frac{1}{n}\sum_{m=0}^{n-1}\sum_{i=0\atop \scriptstyle{\rm even}}^{m+1}\sum_{l=0\atop \scriptstyle{\rm even}}^{n-m}
B_{j}^{(m+1,i)(n-m,l)} \omega_{mi}\omega_{(n-m-1) l},
\ee
which allows us to express the higher coefficients $\omega_{nj}$ through the coefficients with lower values of $n$, starting with $\omega_{00}=1$.
[We remind that $n$ enumerates the loop order and $j\leq (n+{\rm Mod}(n,2))$].
The coefficients   $B_{j}^{(m+1,i)(n-m,l)}$ are given by:
\be
\label{bety}
\nonumber
B_{j}^{(m,i)(p,l)}&=&\frac{1}{2j+1}\left[\frac{N}{2}\delta_{ij}\delta_{lj}+\delta_{ij} \Omega^{li}_{p}
+\delta_{l j} \Omega^{il}_{m}
\right]\\
&+&\left(1+(-1)^j\right)\sum_{k=0}^{{\rm min }[p,m]} \frac{\Omega_m^{ik} \Omega_{p}^{lk} \Omega_{m+p}^{kj}}{2 k+1}\, .
\ee
The recursive relation (\ref{main}) allows a  very fast computation of LLs. For
example, the 33-loop chiral LL is computed in a dozen of seconds
on a PC \footnote{Mathematica notebook for computing LLs is
available at
http://www.tp2.rub.de/$\sim$maximp/research/research.html}.

In present paper we develop the general method for calculation of LL's corrections to the form factors of the Nambu--Goldstone bosons (pions).
We calculate here LL's for the scalar and vector form factors of pions in the massless $O(N+1)/O(N)$ $\sigma$-model (\ref{ON}). This model for $N=3$
is equivalent to the chiral $SU(2)\times SU(2)$ model which describes the leading low-energy
interaction of Nambu--Goldstone bosons (pions) of QCD in the chiral limit \cite{weinberg68}. The chirally odd scalar form factor in the effective
theory (\ref{ON}) is defined as:

\be
\label{scalarFF}
\langle 0|J(0)|\pi^a(p_1)\pi^b(p_2)\rangle =\delta^{ab}\ 2 B F_S(W^2),
\ee
where the scalar form factor $F_S(W^2)$ depends on the invariant mass of two pions $W^2=(p_1+p_2)^2$ and the chirally odd scalar operator $J(x)$
is defined as:

\be
\label{scalarOp}
J(x)= 2 B F \left(F-\sigma(x) \right)=B \sum_{a=1}^N \pi^a(x)\pi^a(x)+O(\pi^4).
\ee
Here the constant $B$ is proportional to the order parameter of spontaneously broken symmetry; in the case of
strong interactions it is proportional to the quark condensate $B=-\langle 0|\bar u u|0\rangle/F^2$.

The chirally even vector form factor in the effective theory (\ref{ON}) is defined as:

\be
\label{vectorFF}
\langle 0| J_\mu^{[ab]}(0)|\pi^c(p_1)\pi^d(p_2)\rangle &=&i(p_2-p_1)_\mu \\
&\times&\left(\delta^{ac}\delta^{bd}-\delta^{bc}\delta^{ad} \right)\ F_V(W^2)\,
\nonumber
\ee
where the vector current $J_\mu^{[ab]}(x)$ is defined as follows:
\be
\label{vectorOP}
J_\mu^{[ab]}(x)=\pi^a(x)\partial_\mu \pi^b(x)-\pi^b(x)\partial_\mu \pi^a(x).
\ee
This current is nothing but the Noether current corresponding to the global $O(N)$
symmetry of the Lagrangian (\ref{ON}).
The low energy expansion of the scalar $F_S(W^2)$ and the vector $F_V(W^2)$ form factors has the following structure:

\be
\label{leFF}
F_{S,V}(W^2)=\sum_{n=0}^\infty \sum_{k=0}^n f^{S,V}_{nk} w^{2n}\ L^k\, ,
\ee
where $f_{nk}^{S,V}$ are the coefficients of the low-energy expansion
in the powers of dimensionless variables
\be
\nonumber
w^2\equiv \frac{W^2}{(4\pi F)^2}, \ \ L\equiv \ln\left(-\frac{\mu^2}{W^2}\right)\, ,
\ee
where $\mu^2$ is the normalization scale.

The lowest coefficients $f_{00}^{S,V}=1$ is obtained from tree-level calculations of the matrix elements (\ref{scalarFF}) and (\ref{vectorFF}).
The calculation of higher order coefficients requires
consideration of the loop diagrams in effective theory. Generically, the coefficients $f_{nk}^{S,V}$ can be obtained
from the calculation of diagrams with number of loops $\leq n$ with inclusion of vertices from higher order effective Lagrangians $\mathcal{L}_{2r}$, which contain
$2r$ derivatives with $r=n+1-k$. The calculation of the LL coefficients $f_n^{S,V}\equiv f_{nn}^{S,V}$ is reduced to calculation
of $n$-loop diagrams with vertices generated  by the the leading effective Lagrangian (\ref{ON}). Presently, record calculations of the LLs coefficients
for the scalar form factor (for $N=3$) are performed by Bissegger and Fuhrer \cite{bissegger} to the four-loop order with the result:
\be
\label{bissegger}
f_1^S=1,\ f_2^S=\frac{43}{36},\ f_3^S=\frac{143}{108},\ f_4^S=\frac{15283}{9720}\, .
\ee
The vector form factor (for $N=3$) is know to the two-loop order \cite{BCT}:
\be
\label{bctresult}
f_1^V=\frac16,\ f_2^V=\frac{1}{72}\,.
\ee

Now we present a general method, which allows us to perform the calculation of LLs for form factors to an unlimited order and for arbitrary $N$. We present
the method for the scalar form factor $F_S(W^2)$, therefore  we do not write (super)subscripts $S$ in order to simplify notations.

The UV divergencies in a $n$-loop diagram are removed by the subtraction of lower-loop graphs with
insertion of the local counterterms corresponding to the subdivergencies of the original $n$-loop diagram. See detailed
discussion of the structure of the subtractions in Refs.~\cite{colangelo,Kivel08}. The local counterterms
relevant
for our calculations renormalize
the couplings $g_{nj}$ of the all-order Lagrangian (see Eq.~(5)of Ref.~\cite{KPVprl}). After subtraction of the UV divergencies the low-energy
expansion of the form factor has the following structure:

\be
\label{ffstruktura}
F(W^2)=\sum_{n=0}^\infty r_n(\mu)\ w^{2n}+\sum_{n=0}^\infty \sum_{k=0}^n f_{nk}({\bf g,r}) w^{2n}\ L^k\, ,
\ee
where $r_n(\mu)$ are renormalized tree level subtraction constants that depend on the renormalization scale $\mu$.
By ${\bf g}$ we denote the infinite set of the constants $g_{nj}(\mu)$ of the all-order chiral Lagrangian (Eq.~(5) of Ref.~\cite{KPVprl}),
by ${\bf r}$ we denote the tree level subtraction constants $r_n(\mu)$.
The form factor $F(W^2)$ should be $\mu$ independent, i.e. $\mu^2 \frac{d}{d\mu^2}F(W^2)=0$. This requirement leads to the following equation
for the coefficients $f_{nk}$:

\be
\label{eq1}
\gamma_n+(\hat{G}+\hat{H}) f_{n0}=0\, ,\\
\nonumber
(\hat{G}+\hat{H}) f_{nk}+(k+1)f_{nk+1}=0\, ,
\ee
where the $\gamma$-functions are defined as $\gamma_n({\bf g,r})\equiv \mu^2 \frac{d}{d\mu^2} r_n(\mu)$ and $\hat{G}$ and $\hat{H}$ stand for the following operators:
\be \label{H}
\hat{G}\equiv \sum_{n=1}^\infty\gamma_n({\bf g,r})\frac{\partial}{\partial r_n}\, ,\ \hat{H}\equiv\sum_{n=1}^\infty \sum_{j=0\atop \scriptstyle{\rm
even}}^{n}\beta_{nj}({\bf g})\frac{\partial}{\partial g_{nj}}\, ,
\ee
with $\beta$-functions defined as $\beta_{nj}({\bf g})\equiv \mu^2 \frac{d}{d\mu^2} g_{nj}(\mu)$. These $\beta$-functions were discussed in details in Ref.~\cite{KPVprl}.

The equation (\ref{eq1}) has an obvious solution: \be \label{solu}
f_{nk}({\bf g,r})=\frac{1}{k!} (\hat{G}+\hat{H})^k \gamma_n({\bf
g,r})\, , \ee with lowest constant $f_{00}=1$ fixed by the tree
order calculation of the form factor with the leading Lagrangian
(\ref{ON}). We see from the solution (\ref{solu}) that, in order
to obtain the LLs (constants $f_n\equiv f_{nn}$), we have to apply
the operator $(\hat G+\hat H)$ $n$ times to the $\gamma$-function
$\gamma_{n}({\bf g,r})$. This at first glance formidable problem
can be solved if one notes that the operators $\hat{G}$ and $\hat
H$ act as a contraction mapping on the space of constants ${\bf
g,r}$. As it was shown in Ref.~\cite{KPVprl} the operator
$\hat H$ possesses the following crucial property:  $\hat{H}^n\
g_{mj}=0,~~ {\rm if}~~m \leq n\,$, which implies that $\hat{H}^n
g_{n+1j}= n!\ \omega_{nj}$ with constants $\omega_{nj}$ satisfying
the non-linear recursion (\ref{main}). Analogously one can show that
the operator $\hat G$ possesses the ``contraction" property $\hat
G^n r_m=0$ if $m\leq n-1$, which implies that the application of the operator $\hat G$
$n$-times to $\gamma_n$ has a fixed point. Thus, we can write:
\be
\label{eqforv}
\hat{G}^n\ \gamma_n({\bf g,r})=\hat{G}^{n+1}\ r_n=n!v_n\, ,
\ee
where $v_n$ are the constants that
are independent of couplings ${\bf g,r}$,
which we use in order to find with help of Eq.~(\ref{solu})
the  coefficients
in front of LLs in form factor expansion $f_n\equiv f_{nn}=v_n$.
For computing the constants $v_n$ (that is equivalent to LL approximation) due to the ``contraction" properties of the
operators $\hat G$ and $\hat H$ only the quadratic part of the
$\gamma_n({\bf g,r})$ contribute to the Eq.~(\ref{eqforv}), so that we can represent the $\gamma$-function as following:

\be
\label{1-loop-gamma}
\gamma_n({\bf g, r})=\sum_{m=0}^{n-1}\sum_{j=0\atop \scriptstyle{\rm even}}^{n-m} \Gamma^{(n-m, j)} r_m g_{(n-m)j}\, .
\ee
The quadratic dependence of the relevant piece of the $\gamma$-function means that the coefficients $\Gamma^{(p,l)}$
can be computed from the one loop diagram shown in Fig.~1.
\begin{figure}
\includegraphics[width =5.cm]{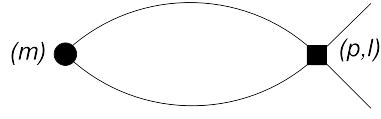}
\caption{Fig.1 One loop diagram contributing to the $\gamma$-functions coefficients (\ref{gammyS},\ref{gammyV}). Filled circle denotes the
counterterms for the form factor and the filled square denotes  counterterms $(pl)$ introduced in Eq.~(5) of Ref.~\cite{KPVprl}. }
\label{fig:halla}
\end{figure}
The result of the calculation for the scalar form factor is:
\be
\label{gammyS}
\Gamma_S^{(p,l)}=\frac{N}{2}\delta_{l0}+ \Omega_p^{l0}\, ,
\ee
and for the vector form factor we obtain:
\be
\label{gammyV}
\Gamma_V^{(p,l)}=(-1)^{p+1}\frac{1}{3}\Omega_p^{l1}\, .
\ee
In both equations matrices $\Omega_p^{lj}$ are given by Eq.~(\ref{omega}).

Now with help of Eq.~(\ref{eqforv}) and Eq.~(\ref{1-loop-gamma}) we can easily obtain recursive equations for the LL coefficients for the scalar and vector
form factors:

\be
\label{mainforf}
f_n^{S,V}=\frac{1}{n}\sum_{m=0}^{n-1}\sum_{j=0\atop \scriptstyle{\rm even}}^{n-m} \Gamma_{S,V}^{(n-m, j)} f_m^{S,V} \omega_{(n-m-1)j},
\ee
with $f_0^{S,V}=1$. The coefficients of $\gamma$-functions  are given by Eq.~(\ref{gammyS}) and Eq.~(\ref{gammyV}) for the scalar and vector form factors correspondingly.
The LL coefficients of the Nambu-Goldstone scattering amplitude $\omega_{pl}$ can be found from the solution of the non-linear recursion relation (\ref{main}).

The equation (\ref{mainforf}) together with (\ref{main}) provides a very powerful tool for calculation of the LL coefficients in the expansion of the form factors.
The results for the first 6 loops for the scalar and vector form factors in the $O(N+1)/O(N)$ sigma-model (\ref{ON}) are presented in Table~1 and Table~2 correspondingly.
For $N=3$ the results coincide with the laborious four-loop calculation of the scalar form-factors obtained in Ref.~\cite{bissegger} (see, Eq.~(\ref{bissegger}))
and with 2-loop calculations of Ref.~\cite{colangelo1}
(see Eq.~(\ref{bctresult}).

In the large $N$ limit the $O(N+1)/O(N)$ sigma model can be solved by the semiclassical methods \cite{Coleman}. In Ref.~\cite{KPVprl} we showed that in this limit the non-linear recursion
equation (\ref{main}) has the following solution :

\be
\omega_{nj}=\left(\frac{N}{2}\right)^n\delta_{j0}\ \left(1+O\left(\frac1N\right)\right)\, .
\ee
Substituting this solution into Eq.~(\ref{mainforf}) we can easily solve corresponding recursion relation in the large $N$ limit with the result ($n\geq 1$):

\be
\label{largeN}
f_n^S&=&\left(\frac{N}{2}\right)^n\ \left(1+O\left(\frac1N\right)\right),\\
\nonumber
f_n^V&=&\frac{1}{(n+1)(n+2)}\left(\frac{N}{2}\right)^{n-1}\ \left(1+O\left(\frac1N\right)\right).
\ee
With these results we can perform the summation of the LL contributions in the large $N$ limit. The form factors in these approximation
have the following form:

\be
F_S(W^2)&=&\frac{1}{1-\varepsilon}\, ,\\
\nonumber
F_V(W^2)&=&1+\frac2N\left(\frac{2-\varepsilon}{2\varepsilon}+\frac{(1-\varepsilon)}{\varepsilon^2}\ln(1-\varepsilon)\right)\, .
\ee
Here we introduced a following short hand notation
$\varepsilon=\frac{N\ W^2}{2 (4\pi F)^2} \ln\left(-\frac{\mu^2}{W^2}\right)$.  We see that the scalar form factor
in the large $N$ limit and LL approximation possesses a pole at $\varepsilon=1$ (which is actually
outside the applicability of the LL approximation). This pole corresponds to the contribution of the auxiliary scalar
field, which is introduced in order to solve the model (\ref{ON}) at $N\to\infty$ (see e.g. \cite{Coleman}). The vector form factor also possesses a weaker
singularity at $\varepsilon=1$, that corresponds to the threshold of $2 \pi+$auxiliary scalar field production.
We see that the form factors in the LL approximation and in the limit of large number of Nambu-Goldstone boson $N$
have correct analytical properties. Let us demonstrate that in general case the form factors obtained in LL
approximation with help of our recursion equation (\ref{mainforf}) satisfy all constraints imposed by
analyticity on the form factors.

Now we present general solution of recursion relation (\ref{mainforf}) that expresses LL coefficients
for form factors $f_n^{S,V}$ in terms of the LL coefficients of $\pi\pi$ scattering amplitude
$\omega_{n,j}$. To this end we introduce two types of generating functions for the coefficients:

\be
\label{generatingff}
{\cal F}_{S,V} (x)&\equiv &\sum_{n=0}^{\infty} f_n^{S,V}\ x^n \, ,\\
\nonumber
W_{S,V}(x)&\equiv&
\sum_{p=1}^{\infty}x^{p-1} \sum_{j=0\atop \scriptstyle{\rm even}}^{p+1} \Gamma^{(p, j)}_{S,V}  \omega_{(p-1)j} \, ,
\ee
which satisfy obvious conditions $W_{S,V}(0)=1$ and ${\cal F}_{S,V}(0)=1$.
The scalar and vector form factors in the LL approximation can be express in terms of generating functions ${\cal F}_{S,V}$
as follows:
\be
\label{relation}
F_{S,V}(W^2)={\cal F}_{S,V}(w^2 L)\, .
\ee
The recursion relation (\ref{mainforf}) can be reduced to the differential equation for the generating functions
(\ref{generatingff}) that has the following solution:

\be \label{solution} {\cal F}_{S,V} (x)=\exp\left(\int_0^x dy\
W_{S,V}(y)\right)\, . \ee
From Eq.~(\ref{AMPpipi}) we can conclude that the lowest $l=0,1$ partial wave amplitudes $t_l^I(s)=\frac{1}{2i}\left(e^{2 i\delta_l^I(s)}-1\right)$ can be expressed in terms of
the generating functions $W_{S,V}(x)$ in the following way:

\be
t_0^0(s)=\pi S W_S(S L), \ \ \ \ t_1^1(s)=\pi S W_V(S L)\, .
\ee
[For notations of kinematical variables see text after Eq.~(\ref{AMPpipi}).]
We see that due to the relations (\ref{relation},\ref{solution}) the form factors in the LL approximation can be expressed in terms
of partial waves for the $\pi\pi$ scattering in the following way:

\be
\nonumber
F_{S(V)}(W^2)&=&\exp\left(\frac{1}{\pi}\int_0^{w^2} \frac{ds}{s}\ t_{0(1)}^{0(1)}(s)\left[\ln\left(-\frac{\mu^2}{s}\right)-1\right]\right)\, .\\
&&\label{pochtiOmnes}
\ee
Noting  the following small $W^2$ asymptotic  to the leading logarithms accuracy:

\be
\nonumber
\lim_{W^2\to 0}\int_0^\infty ds \frac{\left[s\ln\left(-\frac{\mu^2}{s}\right)\right]^n}{s-W^2-i\varepsilon}=\frac1{n+1}W^{2n}\ln^{n+1}\left(-\frac{\mu^2}{W^2}\right)\, ,
\ee
we can put representation of the form factors in terms of the $\pi\pi$ partial wave amplitudes (\ref{pochtiOmnes}) in the form which corresponds to
the Omn\`es solution \cite{Omnes} of the dispersion relations for the form factors:

\be
F_{S(V)}(W^2)=\exp\left(\frac {W^2}\pi\int_0^\infty \frac{ds}{s}\ \frac{\delta_{0(1)}^{0(1)}(s)}{s-W^2-i\varepsilon}\right)\, .
\ee
This demonstrates that the solution of the recursion relation (\ref{mainforf}) provides the form factors with correct analytical properties.

In summary, we suggested new method to compute leading infrared logarithms to arbitrary loop order for the form factors of the
Nambu--Goldstone bosons of an effective field theory in four dimensions. The method is demonstrated on the example of the scalar and the vector
form factors in 4D $O(N+1)/O(N)$ $\sigma$-model.
The proposed method can be applied to a wide range of effective theories.

\section*{Acknowledgements}
Discussions with D.~Diakonov, J.~Gasser, V. Petrov, A.~Rusetsky and O.~Teryaev are greatly appreciated.
 This work was supported in parts by by BMBF,
 by the Deutsche Forschungsgemeinschaft, and by the grant ``Development of Scientific Potential in Higher Schools" (2.2.1.1/1483, 2.1.1/1539).

\newpage
\noindent
Table 1: LL coefficients for the scalar form factor
\begin{equation}\label{tabletildev}
\begin{array}{|l||l|l|}
n & f_n^S(N=3) & f_n^S(N) \\ \hline
 0 & 1 & 1 \\\hline
 1 & 1 & \frac{N}{2}-\frac{1}{2} \\\hline
 2 & \frac{43}{36} & \frac{N^2}{4}-\frac{29 N}{72}+\frac{11}{72} \\\hline
 3 & \frac{143}{108} & \frac{N^3}{8}-\frac{271 N^2}{864}+\frac{7 N}{24}-\frac{89}{864} \\\hline
 4 & \frac{15283}{9720} & \frac{N^4}{16}-\frac{121 N^3}{600}+\frac{423961 N^2}{1555200}-\frac{70997 N}{388800}+\frac{76459}{1555200} \\\hline
 5 & \frac{2578307}{1458000} & \frac{N^5}{32}-\frac{13741 N^4}{108000}+\frac{328547 N^3}{1440000}-\frac{1629803 N^2}{7290000}+\frac{14045881
   N}{116640000}-\frac{169303}{5832000} \\\hline
 6 & \frac{888770227}{428652000} & \frac{N^6}{64}-\frac{6382513 N^5}{84672000}+\frac{3785803199 N^4}{22861440000}-\frac{7206506437
   N^3}{34292160000}+\frac{11173397867 N^2}{68584320000}-\frac{630301337 N}{8573040000}+\frac{255705409}{17146080000}
\end{array}
\nonumber
\end{equation}

\noindent
Table 2: LL coefficients for the vector form factor
\begin{equation}
\begin{array}{|l||l|l|}
n & f_n^V(N=3) & f_n^V(N) \\ \hline
 0 & 1 & 1 \\ \hline
 1 & \frac{1}{6} & \frac{1}{6} \\\hline
 2 & \frac{1}{72} & \frac{N}{24}-\frac{1}{9} \\\hline
 3 & \frac{91}{1296} & \frac{N^2}{80}-\frac{3 N}{160}+\frac{181}{12960} \\\hline
 4 & \frac{3607}{155520} & \frac{N^3}{240}-\frac{17 N^2}{1200}+\frac{27931 N}{1555200}-\frac{4879}{311040} \\\hline
 5 & \frac{7124897}{163296000} & \frac{N^4}{672}-\frac{20917 N^3}{4536000}+\frac{10524683 N^2}{1632960000}-\frac{2081833
   N}{544320000}+\frac{2729}{2551500} \\\hline
 6 & \frac{937784623}{41150592000} & \frac{N^5}{1792}-\frac{24313 N^4}{9408000}+\frac{8249471 N^3}{1524096000}-\frac{478853567
   N^2}{68584320000}+\frac{34922611 N}{6531840000}-\frac{1164883001}{411505920000}
\end{array}
\nonumber
\end{equation}




\begin{thebibliography}{99}
\bibitem{KPVprl}
 N.~Kivel, M.~V.~Polyakov and A.~Vladimirov,
  Phys.\ Rev.\ Lett.\  {\bf 101} (2008) 262001
  [arXiv:0809.3236 [hep-ph]].


\bibitem{weinberg79}
  S.~Weinberg,
  Physica A {\bf 96} (1979) 327.
\bibitem{weinberg68}
  S.~Weinberg,
  Phys.\ Rev.\  {\bf 166} (1968) 1568.
\bibitem{Pagels}
  P.~Langacker, H.~Pagels,
  Phys.\ Rev.\  D {\bf 8} (1973) 4595.
\bibitem{gasser}
  J.~Gasser, H.~Leutwyler,
  Annals Phys.\  {\bf 158} (1984) 142.
\bibitem{colangelo1}
  J.~Bijnens, G.~Colangelo, G.~Ecker, J.~Gasser and M.~E.~Sainio,
  Nucl.\ Phys.\  B {\bf 508} (1997) 263
  [Erratum-ibid.\  B {\bf 517} (1998) 639]
  [arXiv:hep-ph/9707291].
\bibitem{bissegger}
M.~Bissegger and A.~Fuhrer,
  Phys.\ Lett.\  B {\bf 646}, 72 (2007)
  [arXiv:hep-ph/0612096].
\bibitem{BCT}
  J.~Bijnens, G.~Colangelo and P.~Talavera,
  JHEP {\bf 9805} (1998) 014
  [arXiv:hep-ph/9805389].



\bibitem{colangelo}
M.~Buchler and G.~Colangelo,
  Eur.\ Phys.\ J.\  C {\bf 32}, 427 (2003)
  [arXiv:hep-ph/0309049].
\bibitem{Kivel08}
N.~Kivel and M.~V.~Polyakov,
  Phys.\ Lett.\  B {\bf 664} (2008) 64
  [arXiv:0707.2208 [hep-ph]].\\
 N.~Kivel, M.~V.~Polyakov and A.~Vladimirov,
  Phys.\ Rev.\  D {\bf 79} (2009) 014028
  [arXiv:0809.2064 [hep-ph]].

  \bibitem{Coleman}
  S.~R.~Coleman, R.~Jackiw and H.~D.~Politzer,
  Phys.\ Rev.\  D {\bf 10} (1974) 2491.

\bibitem{Omnes}
  R.~Omnes,
  Nuovo Cim.\  {\bf 8} (1958) 316.


\end{thebibliography}
\end{document}